\documentclass[twocolumn,prd,showpacs,nofootinbib,superscriptaddress,
preprintnumbers,floatfix,letterpaper]{revtex4}

\usepackage{amsmath}
\usepackage{amsfonts}
\usepackage{amssymb}
\usepackage{epsfig}
\usepackage{subfigure}

\newcommand{\beq}{\begin{equation}}
\newcommand{\eeq}{\end{equation}}
\newcommand{\beqa}{\begin{eqnarray}}
\newcommand{\eeqa}{\end{eqnarray}}

\begin{document}

\preprint{DESY~07-067, UG-FT-217/07, CAFPE-87/07, ROMA-1452-07}

\title{Long-lived Staus from Cosmic Rays}

\author{Markus Ahlers}
\email{markus.ahlers@desy.de}
\affiliation{Deutsches Elektronen-Synchrotron DESY, 
Notkestra\ss e 85, D-22607 Hamburg, Germany}

\author{Jos\'e Ignacio Illana}
\email{jillana@ugr.es}
\affiliation{CAFPE and
Depto.~de F{\'\i}sica Te\'orica y del Cosmos, Universidad de Granada,
18071 Granada, Spain}

\author{Manuel Masip}
\email{masip@ugr.es}
\affiliation{CAFPE and
Depto.~de F{\'\i}sica Te\'orica y del Cosmos, Universidad de Granada,
18071 Granada, Spain}

\author{Davide Meloni}
\email{meloni@roma1.infn.it}
\affiliation{INFN and 
Dipto.~di Fisica, Universit\`a degli Studi di Roma "La Sapienza", 
00185 Rome, Italy}

\date{May 2007}
\begin{abstract}
  The collision of a high energy cosmic ray with a nucleon in the upper
  atmosphere could produce long-lived heavy particles.  Such particles would
  be very {\it penetrating}, since the energy loss in matter scales as the
  inverse mass, and could reach a neutrino telescope like IceCube from large
  zenith angles.  Here we study this possibility and focus on the long-lived
  stau of SUSY models with a gravitino LSP. The signal would be a pair of
  muon-like parallel tracks separated by $50$ meters along the detector. We
  evaluate the background of muon pairs and show that any events from zenith
  angles above $80^\circ$ could be explained by the production of these heavy
  particles by cosmic rays.
\end{abstract}


\maketitle
\section{Introduction}
The {\it hierarchy problem} has motivated an intense search for new physics
during the past 20 years. Colliders like LEP, the Tevatron, or the B factories
have explored the standard model (SM) at the quantum level but have not
reached the energy or the sensitivity necessary to detect new physics.  There
is experimental evidence for neutrino masses, whereas cosmological data
strongly suggest the presence of a stable weakly-interacting massive particle
(WIMP) as the origin of the dark matter of the universe. However, these
features could be easily {\it added} to the SM without changing its structure.
Therefore, as we approach the search for extensions like supersymmetry (SUSY),
technicolor or extra dimensions at the LHC, it is clear that we should never
underestimate the SM.

On the other hand, cosmic rays are another source of elementary particles of
very high energy with the potential to explore the physics beyond the SM.
When a proton of $10^8$ GeV from outer space hits an atmospheric nucleon it
provides a center of mass energy $\sqrt{s}=\sqrt{2m_N E}$ around $14$~TeV. A
small fraction of these protons (or of the secondary particles with still
enough energy) may then produce exotic massive particles.  Of course, the
question is whether such an event could give any observable signal. In this
paper we argue that this is the case, the long-lived charged particles present
in some extensions of the SM could provide a distinct signature when they
cross a neutrino telescope from large zenith angles.

The process that we propose could take place in SUSY models with an exact
$R$-parity, a gravitino lightest SUSY particle (LSP) working as dark matter,
and a charged next-to-LSP
(NLSP)~\cite{Brandenburg:2005he,Buchmuller:2004rq,Hamaguchi:2004df,
  Feng:2004yi,Hamaguchi:2004ne,Cakir:2007xa,Cyburt:2006uv,Ellis:2006vu,
  Ellis:2003dn,Feng:2004mt}.  The collision of the cosmic proton with the
nucleon could produce any pair of SUSY particles, which would then decay
promptly into the NLSP. Since the NLSP couples very weakly to the LSP
gravitino, it will be long-lived and able to cross a kilometer-long detector
like IceCube.  The generic features of this framework could be also found in
other extensions of the SM, like Little Higgs
models~\cite{Arkani-Hamed:2002qy}.  These models may incorporate a
$T$-parity~\cite{Cheng:2003ju} {\it separating} the standard and the exotic
particles.  The $T$-parity would forbid unobserved mixing between both
sectors, tree-level four fermion operators, and would also make the lightest
particle in the odd sector stable.  If this particle (constituting the dark
matter) is very weakly coupled with the rest, the next-to-lightest one will be
long-lived.

\begin{figure*}[t]
\begin{minipage}[t]{0.48\linewidth}
  \centering\includegraphics[width=\linewidth]{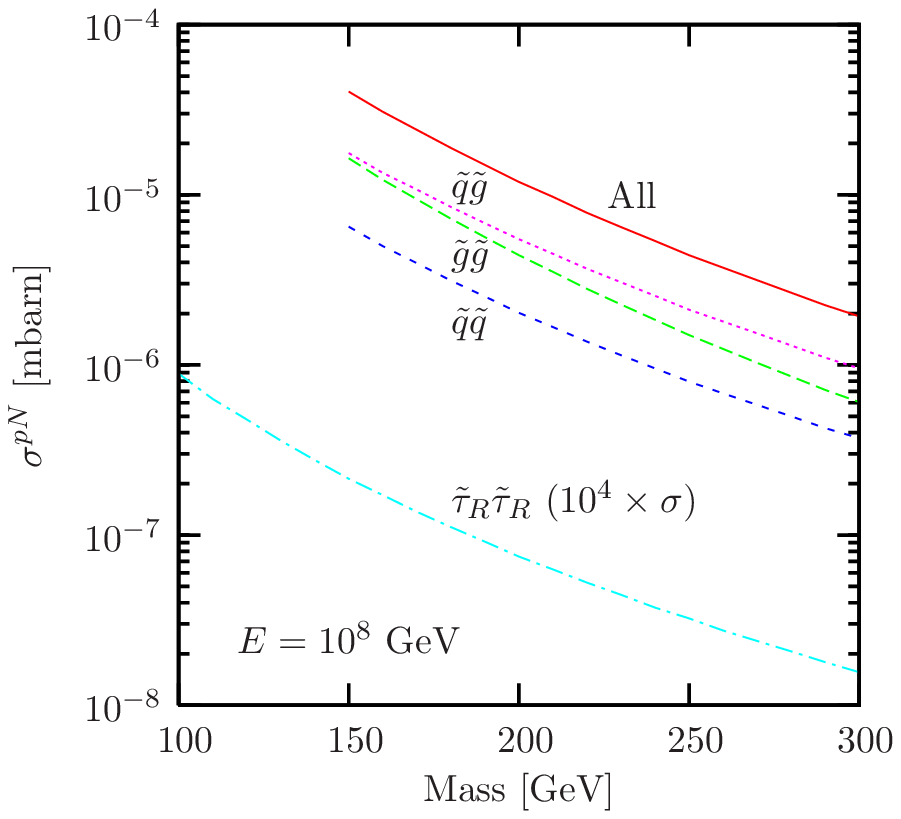}
\caption[]{$pN$ cross section to produce 
  SUSY particles for different values of the stau and the (common)
  squark/gluino mass and a proton energy of $10^8$~GeV.  }\label{fig0}
\end{minipage}
\hfill
\begin{minipage}[t]{0.48\linewidth}
  \centering\includegraphics[width=\linewidth]{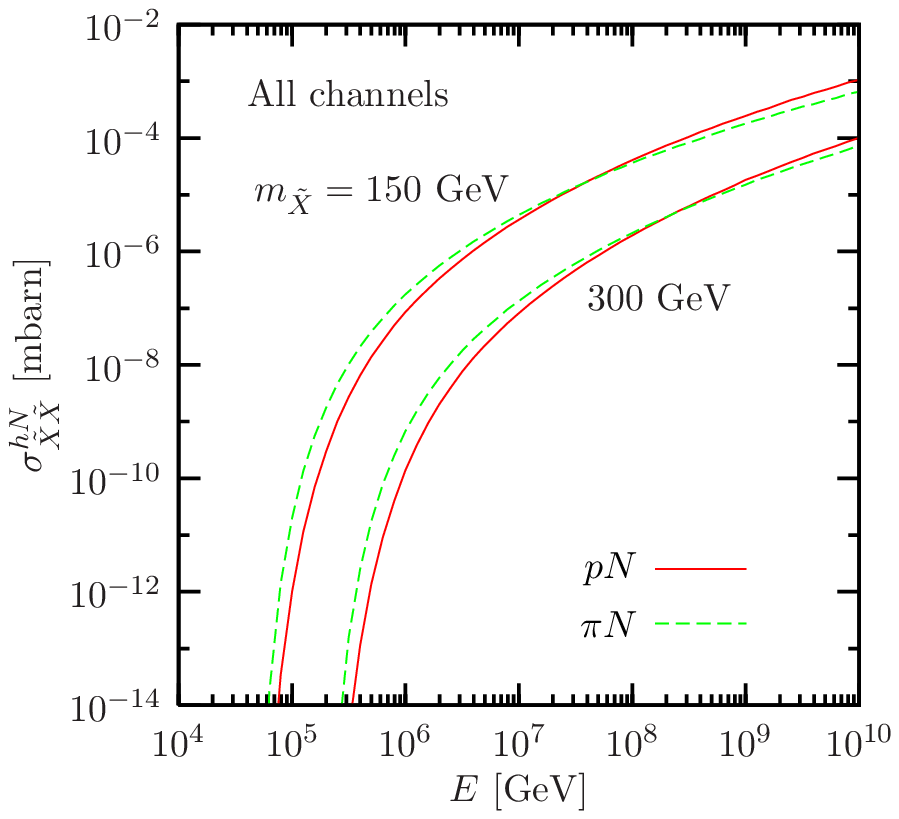}
\caption[]{Cross section to produce any pair of 
  colored SUSY particles in a $pN$ and a $\pi N$ collision for different
  incident energies and a squark/gluino mass $m_{\widetilde X}=150$,
  $300$~GeV.  }\label{fig1}
\end{minipage}
\end{figure*}

The possibility to observe quasi-stable gluinos in
IceCube~\cite{Ahrens:2003ix} has been considered
in~\cite{Hewett:2004nw,Illana:2006xg} in the framework of split-SUSY models
with very heavy sfermions~\cite{Arkani-Hamed:2004fb}.  Here we will focus on
non-colored particles, which present some remarkable differences with the
gluinos. In particular, as one of these particles propagates in matter it
loses energy at a much smaller rate than an
$R$-hadron~\cite{Reno:2005si,Huang:2006ie}.  This fact makes it easier to
confuse it with a muon, but it also lets the particle reach IceCube form
larger zenith angles. To be definite we will consider a long-lived stau
$\widetilde\tau_R$\footnote{We will assume in the following that the
  $\widetilde\tau_R$ life-time is much larger than the propagation time
  through the Earth (see {\it e.g.}~\cite{Giudice:1998bp}).}, although our
arguments would be analogous for any massive charged particle: charginos,
other sleptons, or vectorlike leptons that may appear in Little Higgs models.

Other analyses of the production of exotic particles by cosmic rays refer to
primary
neutrinos~\cite{Albuquerque:2003mi,Bi:2004ys,Ahlers:2006pf,Albuquerque:2006am}.
Being weakly interacting, the relative effect of new physics on the
neutrino-nucleon cross section may be larger (see below), however, it is
difficult to make precise predictions until the flux of cosmic neutrinos is
determined. In contrast, our analysis here relies on a flux of primary cosmic
rays that is well-known in the relevant energy region $10^5$--$10^8$~GeV.

The outline of the paper is as follows. In Sec.~\ref{s1}, we calculate the
rate of long-lived stau pairs produced by collisions of primary and secondary
cosmic rays in the atmosphere.  We discuss then in Sec.~\ref{s2} the signal of
these pairs at IceCube and the background from muon pairs. Section~\ref{s3}
includes a summary of our results.

\begin{figure*}[t]
\begin{minipage}[t]{0.48\linewidth}
  \centering\includegraphics[width=\linewidth]{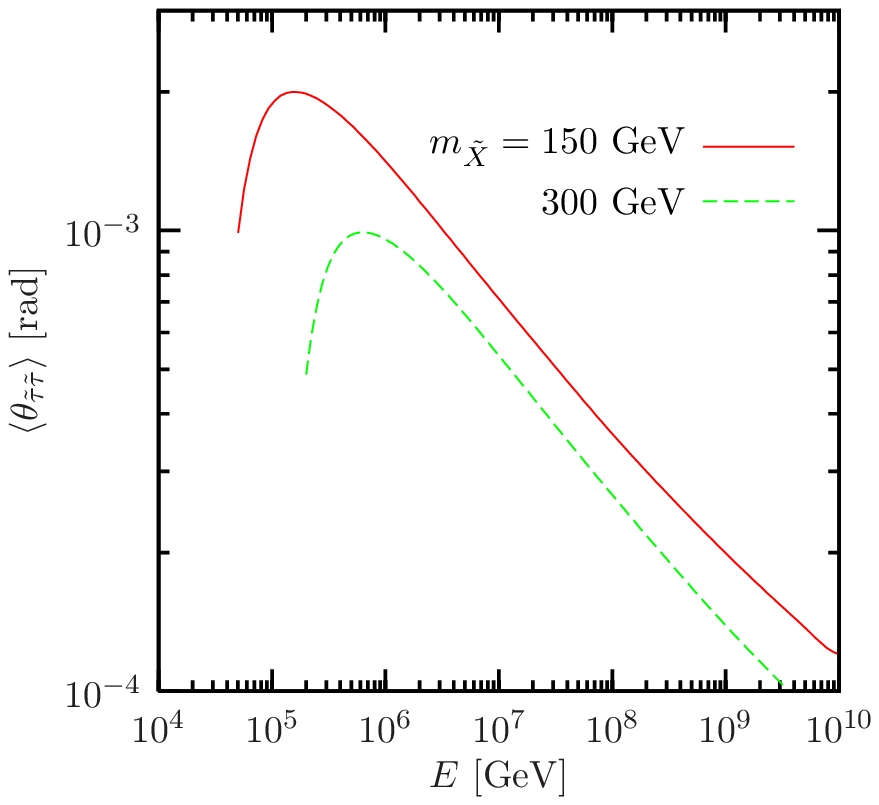}
\caption[]{Average angle between the two 
  staus (in the laboratory frame) for $m_{\widetilde X}=150$, $300$~GeV.}
\label{fig2}
\end{minipage}
\hfill
\begin{minipage}[t]{0.48\linewidth}
  \centering\includegraphics[width=\linewidth]{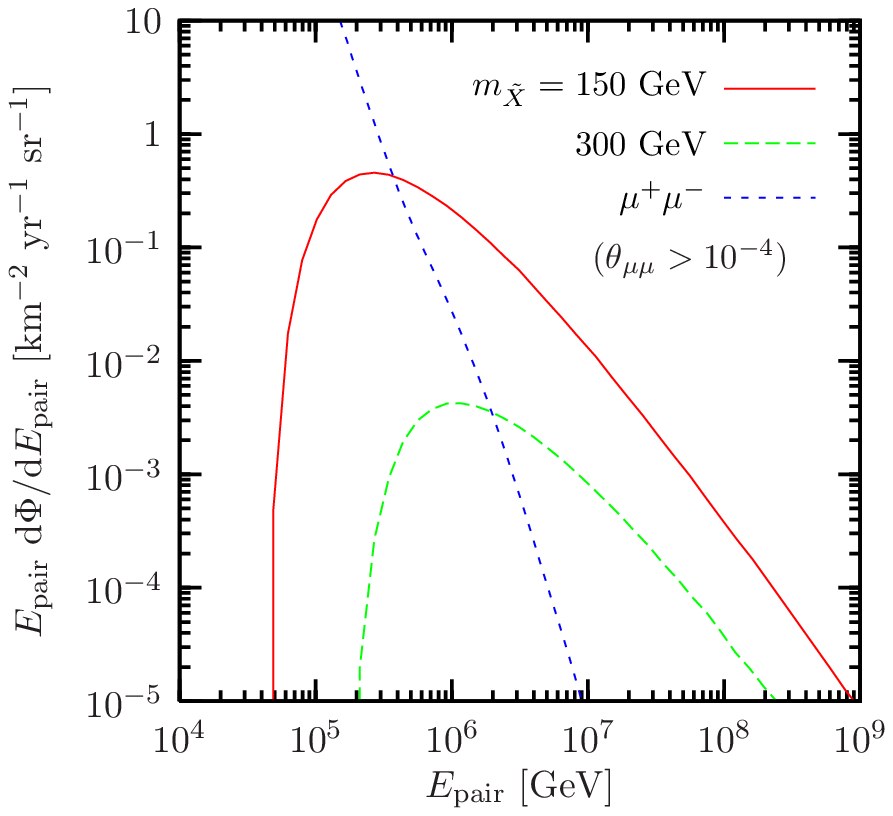}
\caption[]{Flux of stau pairs produced by cosmic rays in terms of the 
  (total) stau energy $E_{\widetilde\tau\widetilde\tau}$ on production for
  $m_{\widetilde X}=150$, $300$~GeV and $\eta=1$. We include the flux of muon
  pairs produced with an opening angle above $10^{-4}$ rad.}\label{fig3}
\end{minipage}
\end{figure*}

\section{Production of stau pairs}\label{s1}

At energies above $10^4$~GeV the decay length of nucleons, charged pions and
kaons is much larger than their interaction length in the air. Therefore, as
they propagate in the atmosphere the probability that one of these hadrons
($h$) collides with a nucleon ($N$) to produce new physics is
just~\cite{Illana:2006mv} \beq {\cal P}^h_X(E) \approx {A \; \sigma^{hN}_X
  \over \sigma^{ha}_{T}}\;.
\label{eq1}
\eeq In this expression $\sigma^{hN}_{X}$ is the cross section to produce the
exotic particle(s) $X$ and $\sigma^{ha}_{T}$ the total cross section of the
hadron with the air (we assume $A=14.6$ nucleons in a nucleus of air and
neglect nuclear effects). The (default) cross sections with the air used by
\mbox{\sc CORSIKA} \cite{Heck:1998vt} above $10^4$~GeV can be approximated by
$\sigma^{ha}_{T}\approx C_0^h + C_1^h \ln (E/{\rm GeV})+C_2^h \ln^2(E/{\rm
  GeV})$, with the constants given in Table~\ref{tab2}. Since
$\sigma^{ha}_{T}$ is above 100 mb, it is apparent that this probability will
be very small and that it would be much larger for a neutrino propagating in
matter.

\begin{table}[b]
\caption{
Constants defining the total cross section with the air 
$\sigma^{ha}_{T}\approx C_0^h +
C_1^h \ln (E/{\rm GeV})+C_2^h \ln^2(E/{\rm GeV})$.
\label{tab2}}
\begin{center}
\begin{tabular}{|c|c|c|c|}
\hline
$h$ & $C_0^h$ [mb] & $C_1^h$ [mb] & $C_2^h$ [mb]
\\
\hline
$N$ & $185.7$ & $13.3$ & $0.08$
\\
$\pi$ & $100.5$ & $16.9$ & $0.00$
\\
$K$ & $79.7$ & $13.9$ & $0.05$
\\
\hline
\end{tabular}
\end{center}
\end{table}

The cross section to produce SUSY particles in a hadronic collision depends
basically on their mass and on whether they have strong interactions. All the
cross sections at the parton level can be found in Ref.~\cite{Dawson:1983fw}.
Collider bounds on SUSY particles with prompt decay into a neutral LSP are
around $250$~GeV for gluinos and squarks, and $100$~GeV for the stop, the
sbottom, charginos, and charged sleptons~\cite{Yao:2006px}. These bounds,
however, may not apply if the particles decay instead into a long-lived
charged or colored SUSY particle.  For example, in order to minimize the SM
background a recent analysis~\cite{Abazov:2006bj} of jets plus missing
momentum at the run II of the Tevatron imposes a veto on events with an
isolated electron or muon with large transverse momentum.  However, gluino or
squark events will include here final staus (instead of neutralinos) that
could be taken by isolated muons. We are not aware of specific bounds on the
colored SUSY spectrum in stau NLSP models.  Notice that bounds based on the
delay in the time of flight versus a muon or the anomalous ionization of the
staus should also be specific, as they are based on the absence of slow-moving
($\beta\le 0.6$) charged particles, but here the staus get an extra boost from
the decay of the parent squark or gluino.  Through the paper we will then
consider slepton, chargino and neutralino masses as low as $100$~GeV and
colored SUSY particles above $150$~GeV.

In Fig.~\ref{fig0} and \ref{fig1} we plot the total $hN$ ($h=p,\; \pi$) cross
sections to produce pairs of these SUSY particles for different SUSY masses
(left panel) and for values of the hadron energy between $10^{4}$ and
$10^{11}$ GeV (right panel).  We have used the {\sc
  CTEQ6M}~\cite{Pumplin:2002vw} ({\sc MRSS}~\cite{Sutton:1991ay}) parton
distribution functions for baryonic (mesonic) interactions, with the
renormalization scale $\mu=0.2 m_{\widetilde X}$ suggested by a NLO
calculation \cite{Beenakker:1996ch}.  We observe that the cross section to
produce {\it directly} a pair of long-lived staus of $100$~GeV is much smaller
than via the production and prompt decay of colored particles of mass around
$200$~GeV. In the latter case, we estimate that the final stau will carry a
fraction $\eta$ \beq \eta \approx {m_{\widetilde X}^2+m_{\widetilde
    \tau}^2\over 2m_{\widetilde X}^2} \;\;\;\;\;(\widetilde X=\widetilde
g,\widetilde q\,)
\label{eta}
\eeq of the energy of the parent gluino or squark. The average angle of the
stau pair (in the lab frame) for different energies of an incident proton are
given in Fig.~\ref{fig2}.

To evaluate the production rate of stau pairs by cosmic rays we need the total
flux of hadrons: primary plus secondary nucleons, pions and kaons produced at
any depth in the atmosphere and with enough energy to create staus in the
collision with an air nucleon.  This analysis has been carried out
in~\cite{Illana:2006xg} assuming a flux of primary nucleons \beq \frac{{\rm
    d}\Phi_{N}}{{\rm d}E}\approx 1.8 \; \left({E\over
    1\,\text{GeV}}\right)^{-2.7}\; {\text{nucleons}\over
  \text{cm}^2\;\text{s}\;\text{sr}\;\text{GeV}}\;,
\label{eq2}
\eeq for energies up to $10^6$ GeV, with a spectral index that changes to 3 in
the interval $10^6$--$10^9$ GeV and goes back to 2.7 at higher energies ({\it
  e.g.}~\cite{Eidelman:2004wy}).  We will use in the following the results on
the flux of secondary hadrons (nucleons, pions, and kaons) derived there.

The flux of quasi-stable staus produced via the prompt decay of a pair
${\widetilde X\widetilde X'}$ of SUSY particles is then \beq \Phi_{\widetilde
  \tau\widetilde \tau} = \sum_{h=N,\pi,K}\; \int^{\infty}_{E_{\rm min}} {\rm
  d}E\;\frac{{\rm d}\Phi_{h}}{{\rm d}E}\; {\cal P}^h_{\widetilde X\widetilde
  X'}(E)\;.
\label{eq4}
\eeq In Fig.~\ref{fig3} we plot the differential flux (d$\Phi$/d$E_{\widetilde
  \tau\widetilde \tau}$) of stau pairs produced by cosmic rays for squarks and
gluinos masses of $150$ and $300$~GeV and $\eta=1$.

\begin{figure*}[t]
\begin{minipage}[t]{0.48\linewidth}
  \centering\includegraphics[width=\linewidth]{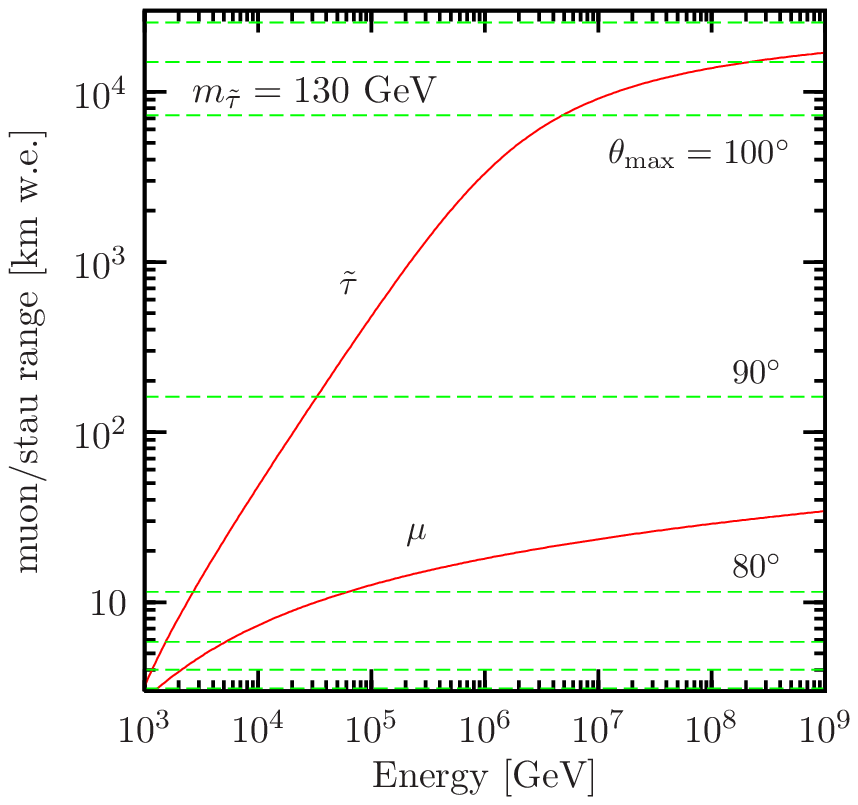}
\caption[]{Range of staus and muons of different energy. 
  The dashed lines show the integrated column depth of the Earth from the
  center of IceCube for increasing zenith angles
  ($\Delta\theta_\text{max}=10^\circ$).}\label{fig4}
\end{minipage}
\hfill
\begin{minipage}[t]{0.48\linewidth}
  \centering\includegraphics[width=\linewidth]{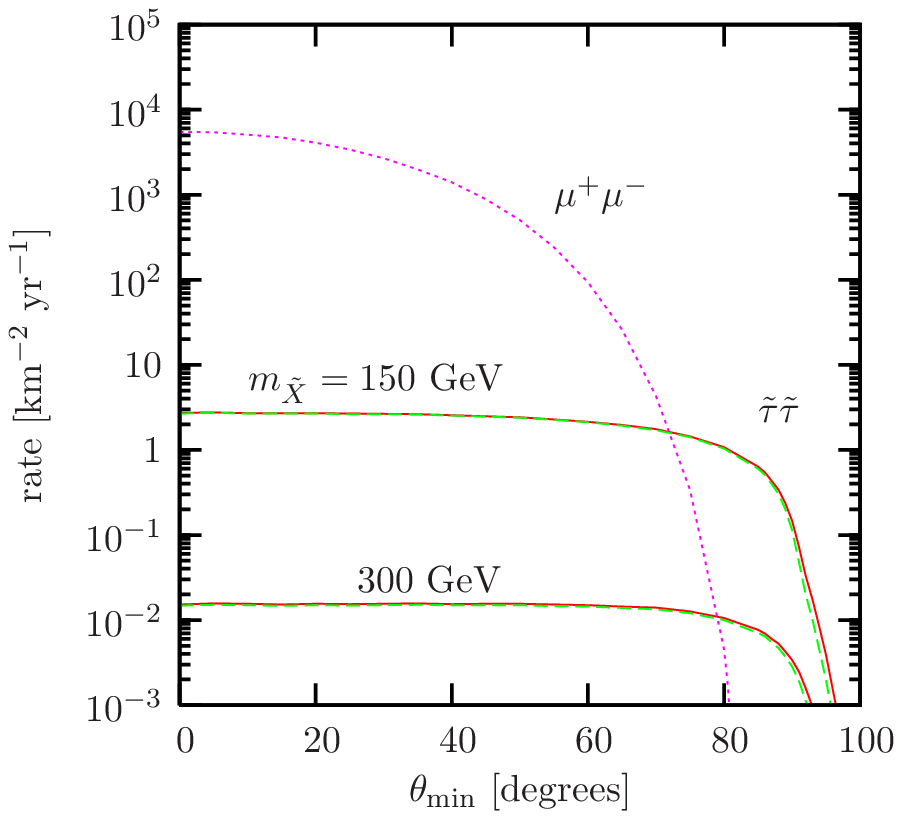}
\caption[]{Integrated number of stau and muon pairs 
  with a minimal separation of $50$ meters at the detector for different
  values of a minimal zenith angle.}\label{fig5}
\end{minipage}
\end{figure*}

\section{Background of muon pairs and signal at IceCube}\label{s2}

The flux of stau pairs produced high in the atmosphere needs to propagate down
to the core of IceCube, about two kilometers under the antarctic ice, to be
observed. In addition, the possible signal faces a strong background of muon
pairs crossing the detector. We plot in Fig.~\ref{fig3} the flux
(d$\Phi$/d$E_{\mu\mu}$) of muon pairs\footnote{We neglect the muons from tau
  decays as they are a $\approx 1\%$ correction to this flux.} produced by
cosmic rays of energy $E_h >10^4$ GeV.  We include only the events where the
two muons are produced with an opening angle above $10^{-4}$ rad, since
smaller angles imply a separation between the two muon tracks that can not be
resolved at IceCube (see below).  Notice that this requirement cuts off muon
pairs with an invariant mass near threshold, $\sqrt{\hat s}\le 1$~GeV, where
the PDFs are mostly unknown and the process would be better described in terms
of hadronic resonances.

The propagation of muons and heavy charged particles in matter is well
understood. For a muon of energy $E_\mu>2 m_\mu$ the mean energy loss per
column density (measured in g/cm$^2$) can be approximated as \beq -{\text{d}
  E_\mu\over \text{d}z}=\alpha_\mu + \beta_\mu E_\mu\;,
\label{enloss}
\eeq where $\alpha_\mu\approx 2\times 10^{-3}$~GeV~cm$^2$/g describes
ionization effects and $\beta_\mu\approx 4\times 10^{-6}$~cm$^2$/g accounts
for bremsstrahlung, pair production and photohadronic processes. The solution
of Eq.~(\ref{enloss}) provides an approximation of the total range of
(initially) very relativistic muons, which we consider in the
following.\footnote{Below $E \approx 2 m$ the ionization energy loss grows
  like $1/\beta^2$ (see {\it e.g.}~\cite{Yao:2006px}).}

For a stau, at the lowest order ionization effects coincide
($\alpha_{\widetilde \tau}\approx \alpha_\mu$) whereas the other effects
depend mainly on the velocity of the particle, which implies
$\beta_{\widetilde \tau}\approx \beta_\mu m_\mu/m_{\widetilde \tau}$.  For
$m_{\widetilde \tau}\approx 100$ GeV, this means that a stau of energy above
$10^5$ GeV losses $10^{3}$ times less energy than a muon of the same energy,
but below $500$ GeV they deposit energy at a similar rate. In our analysis we
will use the approximation for the range of a stau provided
in~\cite{Reno:2005si} and will neglect losses through electroweak
interactions~\cite{Huang:2006ie}, as they are not important for the stau
energies that we obtain.

In Fig.~\ref{fig4} we plot the range of staus and muons of energy between
$10^3$ and $10^{9}$~GeV. We give the correspondence between integrated column
depth of the Earth (see {\it e.g.}~\cite{Gandhi:1995tf}) and zenith angle for
several trajectories ending at the center of IceCube. We observe, for example,
that whereas a muon of $E=10^7$~GeV has a range of about $25$~km water
equivalent (w.e.)  and can reach IceCube from a zenith angle
$\theta_\text{max}\approx 86^\circ$, the range of a stau of the same energy is
$360$ times larger, which makes it able to reach IceCube from zenith angles of
up to $\theta_\text{max}\approx 105^\circ$.

Another relevant observable is the separation of the two particles when they
cross IceCube. This depends on their angle at the creation point (see
Fig.~\ref{fig2}) and the distance from that point to the telescope. The
interaction length of a $10^7$~GeV proton in air is around $4$~g/cm$^2$ (its
cross section is $\sigma^{ha}_{T}\approx 400$~mb), which corresponds to an
altitude of about $20$~km in the atmosphere.  Therefore, if a primary proton
creates a stau pair it will do it around that altitude.  The production of a
stau pair by a secondary hadron will typically occur along the second
interaction length, finishing at around $15$~km, and so on.  To estimate the
distance between the parallel stau tracks at IceCube we will assume that they
are created at a height $H\approx 15$~km. This implies that stau pairs coming
from zenith angles of $60^\circ$, $80^\circ$, and $100^\circ$ fly an
approximate distance of about $30$, $90$, and $2300$~km, respectively, to
reach the center of IceCube.

In Fig.~\ref{fig5} we compare the number of stau and muon pairs reaching
IceCube with a separation larger than $50$ meters, so that the two tracks can
be resolved~\cite{Ribordy:2006qd}.  We plot the flux of these particles coming
from zenith angles larger than the value given in the x-axis ({\it
  e.g.}~$\theta_\text{min}=0$ corresponds to pairs coming from any direction).
We show the cases $m_{\widetilde X}=150$, $300$~GeV and $\eta=1$, $0.7$.  We
observe that from zenith angles between $80^\circ$ and $95^\circ$ there is a
possible signal with no background from muon pairs.

\section{Summary and discussion}\label{s3}

Cosmic rays may be continuously producing massive particles when they collide
with nucleons in the upper atmosphere. If these particles are long-lived they
will be able to reach a detector like IceCube, about two kilometers under the
ice.  To have a sizeable production rate (order $1$ per year and square
kilometer) the particles should be produced through strong interactions. This
would be the case for a quasi-stable stau resulting from the prompt decay of a
gluino or a squark.  We have studied in some detail the possibility to observe
such an event.

The heavy staus would be produced in pairs at altitudes around $15$~km, and as
they approach IceCube the two staus would separate.  In principle, they could
be confused with a muon pair: a muon and a stau of $500$~GeV would give in
IceCube a very similar signature.  We have shown, however, that it is possible
to reduce the dimuon background below the signal.  Above $E\approx 500$~GeV
muons lose energy in ice much faster than the staus. As a consequence, while
muons will never reach IceCube from directions close to the horizon, staus can
come from zenith angles of up to $110^\circ$.  In addition, larger zenith
angles mean also larger distance of flight and, in turn, larger separation
between the two tracks at IceCube.

We obtain that any events with two muon-like tracks separated by more than
$50$ meters coming from zenith angles of $80^\circ$--$100^\circ$ would be a
clear signal of heavy charged particles produced by cosmic rays high in the
atmosphere (see Fig.~\ref{fig5}).

One may wonder if this signal could also be distinguished from possible stau
pairs produced by primary neutrinos, which have been extensively considered in
the literature.  We find two main differences with such an event. Suppose that
the primary comes horizontally ($\theta=90^\circ$). In our case the staus will
be created high in the atmosphere, whereas in a neutrino event the interaction
to produce them will typically take place deep inside the ice. This implies a
smaller distance between the two stau tracks along IceCube.  A second
important difference is that while neutrino events could come basically from
any direction, the staus produced by primary protons vanish at zenith angles
above about $115^\circ$.

We think that the study of inclined ($\theta\ge 60^\circ$) two-muon events at
IceCube would be interesting by itself. For example, it can be used as an
indirect measure of the total (primary plus secondary) flux of hadrons of
$10^4$--$10^9$~GeV and, thus, as a test for the different codes that simulate
extensive air showers. Going to zenith angles below the horizon the SM
background vanishes, leaving some room for exotic physics.

\section*{Acknowledgments}

We would like to thank Francisco del \'Aguila, 
Fernando Cornet, Andreas Ringwald, and Christian Spiering for 
useful discussions. This work has been supported by MEC of Spain 
(FPA2006-05294) and Junta de Andaluc\'\i a 
(FQM-101). We also acknowledge financial support from a 
grant CICYT-INFN (07-10).


\bibliographystyle{h-physrev3}
\frenchspacing
\bibliography{refs}


\end{document}